\documentclass[twocolumn,showpacs,preprintnumbers,superscriptaddress,aps]{revtex4-1}

\usepackage{graphicx,color}

\usepackage{dcolumn}

\usepackage{bm}
\usepackage{here}
\usepackage[]{lineno}
\usepackage{mathrsfs} 

\begin{document}

\preprint{}

\title{Large spontaneous Hall effect with flexible domain control \\
in an antiferromagnetic material TaMnP}

\author{Hisashi Kotegawa}
\affiliation{Department of Physics, Kobe University, Kobe, Hyogo 657-8501, Japan}

\author{Akira Nakamura}
\affiliation{Department of Physics, Kobe University, Kobe, Hyogo 657-8501, Japan}

\author{Vu Thi Ngoc Huyen}
\affiliation{Center for Computational Materials Science, Institute for Materials Research, Tohoku University, Sendai, Miyagi 980-8577, Japan}

\author{Yuki Arai}
\affiliation{Department of Physics, Kobe University, Kobe, Hyogo 657-8501, Japan}

\author{Hideki Tou}
\affiliation{Department of Physics, Kobe University, Kobe, Hyogo 657-8501, Japan}

\author{Hitoshi Sugawara}
\affiliation{Department of Physics, Kobe University, Kobe, Hyogo 657-8501, Japan}

\author{Junichi Hayashi}
\affiliation{Muroran Institute of Technology, Muroran, Hokkaido 050-8585, Japan}

\author{Keiki Takeda}
\affiliation{Muroran Institute of Technology, Muroran, Hokkaido 050-8585, Japan}

\author{Chihiro Tabata}
\affiliation{Materials Sciences Research Center, Japan Atomic Energy Agency, Tokai 319-1195, Japan}
\affiliation{Advanced Science Research Center, Japan Atomic Energy Agency, Tokai 319-1195, Japan}

\author{Koji Kaneko}
\affiliation{Materials Sciences Research Center, Japan Atomic Energy Agency, Tokai 319-1195, Japan}
\affiliation{Advanced Science Research Center, Japan Atomic Energy Agency, Tokai 319-1195, Japan}

\author{Katsuaki Kodama}
\affiliation{Materials Sciences Research Center, Japan Atomic Energy Agency, Tokai 319-1195, Japan}

\author{Michi-To Suzuki}
\affiliation{Center for Computational Materials Science, Institute for Materials Research, Tohoku University, Sendai, Miyagi 980-8577, Japan}
\affiliation{Center for Spintronics Research Network, Graduate School of Engineering Science, Osaka University, Toyonaka, Osaka 560-8531, Japan
}

\date{\today}

\begin{abstract}
Antiferromagnets without parity-time ($\mathcal{PT}$) symmetry offer novel perspectives in the field of functional magnetic materials.
Among them, those with ferromagnetic-like responses are promising candidates for future applications such as antiferromagnetic (AF) memory; however, examples showing large effects are extremely limited.
In this study, we show that the orthorhombic system TaMnP exhibits a large anomalous Hall conductivity (AHC) $\sim360-370$ $\Omega^{-1}$cm$^{-1}$ in spite of the small net magnetization $\sim10^{-2}$ $\mu_B$/Mn.
Our neutron scattering experiment and the observation of the AH effect indicated that a magnetic structure of TaMnP was dominated by an AF component represented by $B_{3g}$ with the propagation vector $q=0$.
Furthermore, we confirmed that the obtained AHC is among the largest observed in AF materials at zero fields. 
Additionally, our first-principles calculations revealed that the spin-orbit interaction originating in the nonmagnetic Ta-$5d$ electrons significantly contributes to enhancing Berry curvatures in the momentum space.
We found that the magnetic fields along all the crystal axes triggered the AF domain switching, indicating the possibility of controlling the AF domain using the small net magnetization, which is symmetrically different.

\end{abstract}

\maketitle

\section{Introduction}

Conventionally, the industrial value of antiferromagnets without net magnetization is significantly lower than that of ferromagnets owing to their lack of magnetic field responses.
However, the controllability of antiferromagnetic (AF) domains has been recently developed from several perspectives.
One direction involves the application of antiferromagnets without $\mathcal{PT}$ symmetry ($\mathcal{P}$: space-inversion symmetry, $\mathcal{T}$: time-reversal symmetry).
In conventional antiferromagnets, opposite-spin sublattices, which are connected by translation or inversion symmetry, protect the compensated total magnetic moment.
The latter corresponds to the $\mathcal{PT}$ symmetry, which guarantees compensation even at a zero propagation vector. 
When the $\mathcal{PT}$ symmetry is broken, an AF spin configuration could be represented by the magnetic point group (MPG) that allows the existence of a ferromagnetic (FM) state.
In this case, nonzero net magnetization can be induced from the symmetry.
As net magnetization is coupled to the AF structure, magnetic fields can flip the AF domain.
Another benefit brought by the symmetry is that the AF structure notably offers FM-like functional responses \cite{Chen14,Suzuki17,Smejkal20}, and the magnetic fields are used to change the sign of the responses.
The typical response is the anomalous Hall effect (AHE) at zero field \cite{Nakatsuji2015,Kiyohara16,Nayak16,Ghimire18,Park22}, which is a spontaneous Hall effect that had been believed to stem from the large net magnetization of ferromagnets \cite{Hall,Manyala2004,Husmann2006}.
These kinds of AF structures also yield other FM-like responses, such as anomalous Nernst effect, magneto-optical Kerr effect, and magnetic spin Hall effect \cite{Li17,Ikhlas17,Higo18,Kimata19}.
A magnetic symmetry equivalent to FM states allows nonzero anomalous Hall conductivity (AHC); however, the numbers of up- and down-spin conduction electrons in antiferromagnets are the same. Therefore, the crucial ingredients for significant responses are summarized in the electronic band structure, which generates the imbalanced distribution of the large Berry curvatures between positive and negative values under symmetry breaking in the momentum space.
The highest AHC arising from the AF structure, which is observable at zero fields and ambient pressure, is 380 $\Omega^{-1}$cm$^{-1}$ in Mn$_3$Ge \cite{Kiyohara16}.
This value is comparable with that of ferromagnets ($\sim10^2-10^3$ $\Omega^{-1}$cm$^{-1}$) \cite{Nagaosa10}, and other AF materials exhibiting such large responses have not been reported.
Notably, the contribution of Weyl points to the origin of such high AHC of Mn$_3$Ge has been reported, indicating the existence of a singularity of the electronic state for this material \cite{Chen2021}.
Consequently, it remains an open question whether such high AHC is obtained under extremely restricted conditions or whether it can be generally observed in other AF materials.

Recently, we reported the emergence of large AHE in a noncollinear AF material NbMnP \cite{Kotegawa_NbMnP}.
The AF structure with $q=0$ was expressed by a linear combination of magnetic structures belonging to irreducible representations (IRs), $B_{2u}$ and $B_{3g}$ \cite{Matsuda}.
The $B_{3g}$ representation corresponds to a magnetic space group $Pnm'a'$ (a MPG $mm'm'$).
This AF structure resulted in the same symmetry breaking as the FM structure along the $a$-axis, thereby confirming that it yielded a nonzero AHE.
The experimental AHC value ($230$ $\Omega^{-1}$cm$^{-1}$) was almost reproduced theoretically using Berry curvatures summed over a momentum space, demonstrating that the AHE stems from the AF spin configuration \cite{Kotegawa_NbMnP}.
Furthermore, the AHC value ($230$ $\Omega^{-1}$cm$^{-1}$) was $60$ \% of Mn$_3$Ge, indicating the sufficient potential of this crystal and magnetic structure.
Here, we explored another AF material TaMnP, which is isostructural to NbMnP and whose physical properties have not been clarified \cite{NbMnP,TaMnP}.

For the exploration, we synthesized TaMnP single crystals via a Ga-flux method for the first time.
The crystals exhibited an AF transition at $220\sim240$ K, which was accompanied by a small net magnetization ($\sim10^{-2} \mu_{\rm B}$/Mn).
In the ordered state, high AHC ($\sim360-370$ $\Omega^{-1}$cm$^{-1}$) was observed, representing one of the highest AHCs from AF structures.
Notably, the small net magnetization was observed, even along a different direction to that in NbMnP, comfirming the flexible controllability of the AF domain in TaMnP.

\section{Methods}

\subsection{Sample Preparation}
To grow the single crystals, the starting material comprising Ta powder, Mn flakes, P flakes, and Ga grains (molar ratio: 1:10:1:30), was placed in an Al$_2$O$_3$ crucible and sealed in an evacuated quartz ampoule. 
Excess Mn was required to avoid the growth of TaP.
Next, we gradually heated the ampoule to 1050 ${\rm ^o C}$ for 6 h, followed by slow cooling to 750 ${\rm ^o C}$ at $-5$ ${\rm ^o C}$/h.
After centrifugation, the excess substances were dissolved in an acid solution, yielding needle-like single crystals along the $b$-axis.
The crystal symmetry, lattice parameters, and occupancy of each site were assessed by single-crystal X-ray diffraction (XRD) measurements using a Rigaku Saturn724 diffractometer.

\subsection{Electrical and Magnetization Measurements}

To measure electrical resistivity and Hall resistivity, we made electrical contacts with gold wires using the spot-weld method. 
Both quantities were measured using a standard four-probe method. We furthermore antisymmetrized the Hall resistivity against magnetic fields to remove the longitudinal component induced by potential contact misalignments. 
The magnetization measurements were performed using a Magnetic Property Measurement System (Quantum Design), where several pieces of aligned single crystals (approximately 20 crystals) were fixed against magnetic fields.

\subsection{Neutron Scattering Measurements}
Two sets of neutron powder diffraction experiments were performed using the research reactor JRR-3 in Tokai, Japan. 
First, powder diffraction patterns were collected using a high-resolution powder diffractometer HRPD at representative temperatures (13.5 and 260 K) to determine the structure with a wavelength of $\lambda$~=~1.823~\AA~ monochromatized with a Ge(3 3 1) monochromator and a collimation of open-12'-6'. 
The resulting powder diffraction patterns were analyzed using the FullProf software \cite{FullProf}.
Next, the temperature variation of a magnetic peak was measured using the triple-axis spectrometer (TAS-1). 
A pyrolytic graphite (PG) monochromator and analyzer crystals were employed with a collimation of open-80'-80'-80' at $E_i$=14.7~meV together with PG filters to reduce the higher-order contamination.
The sample was sealed in a vanadium cell and cooled in a GM-type closed-cycle refrigerator and $^4$He-based top-load cryostat \cite{Kaneko} on HRPD and TAS-1, respectively.

\vspace{30ex}

\subsection{First-Principles Calculation}

The first-principles calculations were performed using the QUANTUM ESPRESSO package \cite{Giannozzi}. 
The generalized gradient approximation (GGA) in the parametrization of Perdew, Burke, and Ernzerhof \cite{Perdew} was employed for the exchange-correlation functional, and the pseudopotentials in the projector augmented-wave method \cite{Blochl,Kresse} were generated by PSLIBRARY \cite{Corso}. 
Beginning from the experimentally obtained atomic positions in this study the atomic positions were fully relaxed until residual forces $< 0.01$ eV/\AA\ were reached. 
Next, we adopted kinetic cutoff energies of 50 and 400 Ry as the plane-wave basis set and charge density, respectively. 
To obtain the Fermi level, a $k$ mesh of $9\times15\times9$ was used to sample the first Brillouin zone (BZ) with a Methfessel- Paxton smearing width of 0.005 Ry.
Furthermore, we used the PAOFLOW package~\cite{Naredelli,Cerasoli} for AHC estimation. 
An 18 $\times$ 30 $\times$ 18 Monkhorst-Pack $k$-point grid was used to generate a tight-binding set of pseudoatomic orbitals for the subsequent AHC calculations.
Wannier90 \cite{Giovanni} was used to plot the Berry curvature in BZ. 
The $4d$ and $5s$ orbitals of Nb, $5d$ and $6s$ orbitals of Ta, $3d$ and $4s$ orbitals of Mn, and $3s$ and $3p$ orbitals of P were included in the Wannier interpolation scheme using Wannier90 to construct realistic tight-binding models from the first-principles band structures \cite{Wang}. 

AHC was calculated using the Kubo formula \cite{Wang}.
\begin{equation}
\sigma_{\alpha \beta} = -\frac{e^2}{\hbar} \int \frac{d \bm{k}}{(2 \pi)^3} \sum_n f [ \varepsilon_n(\bm{k})-\mu ] \Omega_{n,\alpha \beta}(\bm{k}),
\end{equation}
where $n$ is the band index and $\alpha, \beta = x, y,$ and $z$ with $\alpha \neq \beta$.
The Berry curvature was defined, as follows:
\begin{equation}
\Omega_{n,\alpha \beta}(\bm{k}) = - 2 {\rm Im} \sum_{m\neq n} \frac{v_{nm,\alpha}(\bm{k}) v_{mn,\beta}(\bm{k}) }{[ \varepsilon_m(\bm{k})-\varepsilon_n(\bm{k}) ]^2}.
\end{equation}
Here, $\varepsilon_n(\bm{k})$ is the eigenvalue, and 
\begin{equation}
v_{nm,\alpha}(\bm{k}) = \frac{1}{\hbar} \left< u_n(\bm{k}) \left| \frac{\partial \hat{H}(\bm{k})}{\partial k_{\alpha}} \right| u_m(\bm{k}) \right>,
\end{equation}
where $u_n(\bm{k})$ is the periodic cell part of the Bloch states and $\hat{H}(\bm{k}) = e^{-ik \cdot r} \hat{H} e^{ik \cdot r}$.

\section{Results and Discussion}

\subsection{Crystal Structure}

The results of single-crystal XRD measurements of the obtained TaMnP crystal are summarized in Table I.
The lattice parameters were almost consistent with those reported for a polycrystalline sample \cite{TaMnP} and similar to those of NbMnP \cite{Matsuda}.
Notably, the Ta and Mn sites exhibited deficiencies of $4-5$ \%.
Figure 1 shows the crystal structure, revealing that nonsymmorphic TaMnP comprised four equivalent Mn atoms in a unit cell.
Additionally, the second-, third-, and fourth-nearest neighbor Mn-Mn bonds lack inversion symmetry.

\begin{table}[htb]
\caption{Structural parameters of TaMnP determined by single-crystal XRD measurements at $T=293$ K. The Wyckoff positions of all the atoms are $4c$. The data for NbMnP, which was synthesized by a self-flux method, are shown for reference \cite{Matsuda}. }
\label{t1}
\raggedright
\begin{center}
\begin{tabular}{ccccccc}
\hline
TaMnP & 293 K & & & & \\
\hline
Atom & $x$ & $y$ & $z$ & Occ. &$U$(\AA$^2$)\\
\hline
Ta & 0.02981(8) & 0.25000 & 0.67250(6) & 0.952 & 0.0048(3)\\
Mn & 0.1438(3) & 0.25000 & 0.0600(2) & 0.963 & 0.0056(6)\\
P & 0.2704(5) & 0.25000 & 0.3701(4) & 1 & 0.0051(9)\\
\hline
\end{tabular}\\
orthorhombic ($Pnma$) \\
$a$=6.1323(2) \AA, $b$=3.55690(10) \AA, $c$=7.1669(3) \AA, $R$=2.70\% \\
\vspace{5ex}
\begin{tabular}{ccccccc}
\hline
NbMnP & 293 K & & & & \\
\hline
Atom & $x$ & $y$ & $z$ & Occ. &$U$(\AA$^2$)\\
\hline
Nb & 0.03102(5) & 0.25000 & 0.67215(5) & 0.968 & 0.00582(15)\\
Mn & 0.14147(9) & 0.25000 & 0.05925(8) & 1 & 0.0063(2)\\
P & 0.26798(15) & 0.25000 & 0.36994(13) & 1 & 0.0061(2)\\
\hline
\end{tabular}\\
orthorhombic ($Pnma$) \\
$a$=6.1823(2) \AA, $b$=3.5573(2) \AA, $c$=7.2187(3) \AA, $R$=1.90\% \\
\end{center}
\end{table}

\begin{figure}[htb]
\includegraphics[width=8cm]{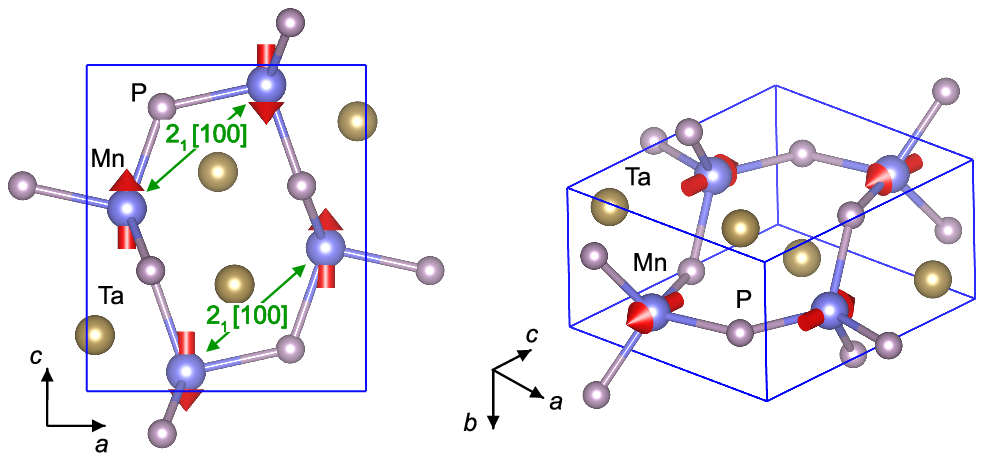}
\caption{Crystal structure of TaMnP in the orthorhombic $Pnma$ space group. A suggested $\mathcal{PT}$ symmetry-broken magnetic structure, represented by $B_{3g}$, is shown. The unit cell (blue lines) includes four Mn atoms. The AF coupling between two Mn atoms is connected by a screw operation, $2_1$, along the $a$-axis. The screw axis is $(x, 1/4, 1/4)$. Lack of the $\mathcal{PT}$ symmetry does not protect compensation of the total magnetic moment, inducing the FM component along the $a$ axis through DM interaction.}
\label{structure}
\end{figure}

\subsection{Electrical Resistivity and Magnetization}

\begin{figure}[h]
\includegraphics[width=8.6cm]{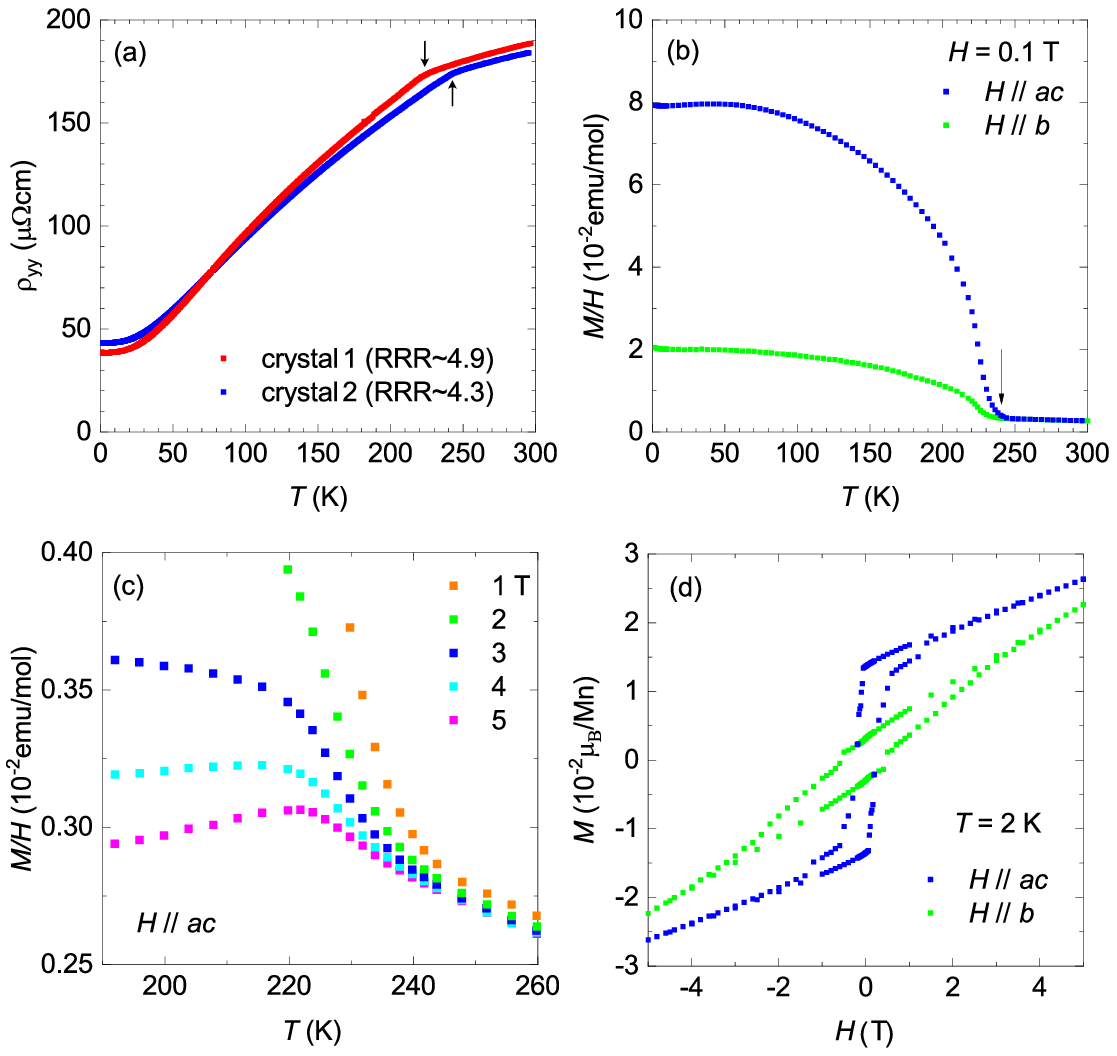}
\caption{(a) Electrical resistivity $\rho_{yy}$ and (b-d) magnetization of TaMaP. The magnetic transition was observed at $220-240$ K with nonnegligible sample dependence. The magnetization was measured for several pieces of aligned single crystals. The kinks in $M/H$, which appeared under high magnetic fields, and the small net magnetization of $\sim 1.4 \times 10^{-2} \mu_B$/Mn indicate that the AF component dominated the magnetic structure of TaMnP. }
\end{figure}

\begin{table*}[htb]
\caption{Allowed magnetic moments at four Mn sites in a unit cell for each IR in $Pnma$. The magnetic propagation vector of (0, 0, 0) was assumed. $u$, $v$ and $w$ represent magnetic moment components. The corresponding MPGs are shown. The magnetic structure in NbMnP is represented by a combination of $B_{2u}$ and $B_{3g}$ \cite{Matsuda}. The $B_{1g}$, $B_{2g}$, and $B_{3g}$ allow the spontaneous net magnetization and nonzero AHC.}
\label{t2}
\raggedright
\begin{center}
\begin{tabular}{>{\centering}p{4em}>{\centering}p{8em}>{\centering}p{10em}>{\centering}p{10em}>{\centering}p{12em}>{\centering}p{5em}c}
\hline
IR & Mn1 & Mn2 & Mn3 & Mn4 & MPG & AHC\\
& ($x, y, z$) & ($-x$+1/2, $-y$, $z$+1/2) & ($-x$, $y$+1/2, $-z$) & ($x$+1/2, $-y$+1/2, $-z$+1/2) & &\\
\hline
$A_{g}$ ($\Gamma_1$) & (0, $v$, 0) & (0, $-v$, 0) & (0, $v$, 0) & (0, $-v$, 0) & $mmm$ & \\
$A_{u}$ ($\Gamma_2$) & ($u$, 0, $w$) & ($-u$, 0, $w$) & ($-u$, 0, $-w$) & ($u$, 0, $-w$) & $m'm'm'$ &\\
$B_{1g}$ ($\Gamma_3$) & ($u$, 0, $w$) & ($-u$, 0, $w$) & ($u$, 0, $w$) & ($-u$, 0, $w$) & $m'm'm$ & $\sigma_{xy}\neq0$ \\
$B_{1u}$ ($\Gamma_4$) & (0, $v$, 0) & (0, $-v$, 0) & (0, $-v$, 0) & (0, $v$, 0) & $mmm'$ &\\
$B_{2g}$ ($\Gamma_5$) & (0, $v$, 0) & (0, $v$, 0) & (0, $v$, 0) & (0, $v$, 0) & $m'mm'$ & $\sigma_{zx}\neq0$ \\
$B_{2u}$ ($\Gamma_6$) & ($u$, 0, $w$) & ($u$, 0, $-w$) & ($-u$, 0, $-w$) & ($-u$, 0, $w$) & $mm'm$ &\\
$B_{3g}$ ($\Gamma_7$) & ($u$, 0, $w$) & ($u$, 0, $-w$) & ($u$, 0, $w$) & ($u$, 0, $-w$) & $mm'm'$ & $\sigma_{yz}\neq0$ \\
$B_{3u}$ ($\Gamma_8$) & (0, $v$, 0) & (0, $v$, 0) & (0, $-v$, 0) & (0, $-v$, 0) & $m'mm$ &\\
\hline
\end{tabular}\\
\end{center}
\end{table*}

Figures 2 (a-d) show the electrical resistivity ($\rho_{yy}$) and magnetization of TaMnP.
Here, the $a$-, $b$-, and $c$-axes correspond to $x$, $y$, and $z$, respectively.
These figures reveal a magnetic transition in TaMnP at approximately $T_{\rm N} = 220-240$ K.
The kinks in $\rho_{yy}$ appeared at 225 K for crystal~1 and 242 K for crystal~2, indicating the distribution of the transition temperatures among crystals, probably owing to the deficiencies of the Ta or Mn atoms (Table I). 
The residual resistivity ratios (RRRs) of crystal~1 and crystal~2 was 4.9 and 4.3, respectively, which is higher than that of NbMnP (RRR$\sim2$) \cite{Matsuda,Kotegawa_NbMnP}.
We also observed an increase in the measured magnetization at 0.1 T below $230-240$ K (Fig.~2(b)). 
It appeared like an FM transition, whereas the kinks were observed under higher magnetic fields, similar to those of NbMnP exhibiting an AF structure \cite{Matsuda} (Fig.~2(c)).
This indicated that the AF component dominated the magnetic structure of TaMnP. 
Notably, no large spontaneous magnetization was observed even at 2 K (Fig.~2(d)), and its magnitude was at most $1.4\times10^{-2} \mu_B$/Mn for the $H \parallel ac$ plane.
The presence of small net magnetization indicated that the propagation vector of TaMnP is $q=0$.

\subsection{Neutron Scattering Measurements and Symmetrical Consideration}

\begin{figure}[htb]
\includegraphics[width=9cm]{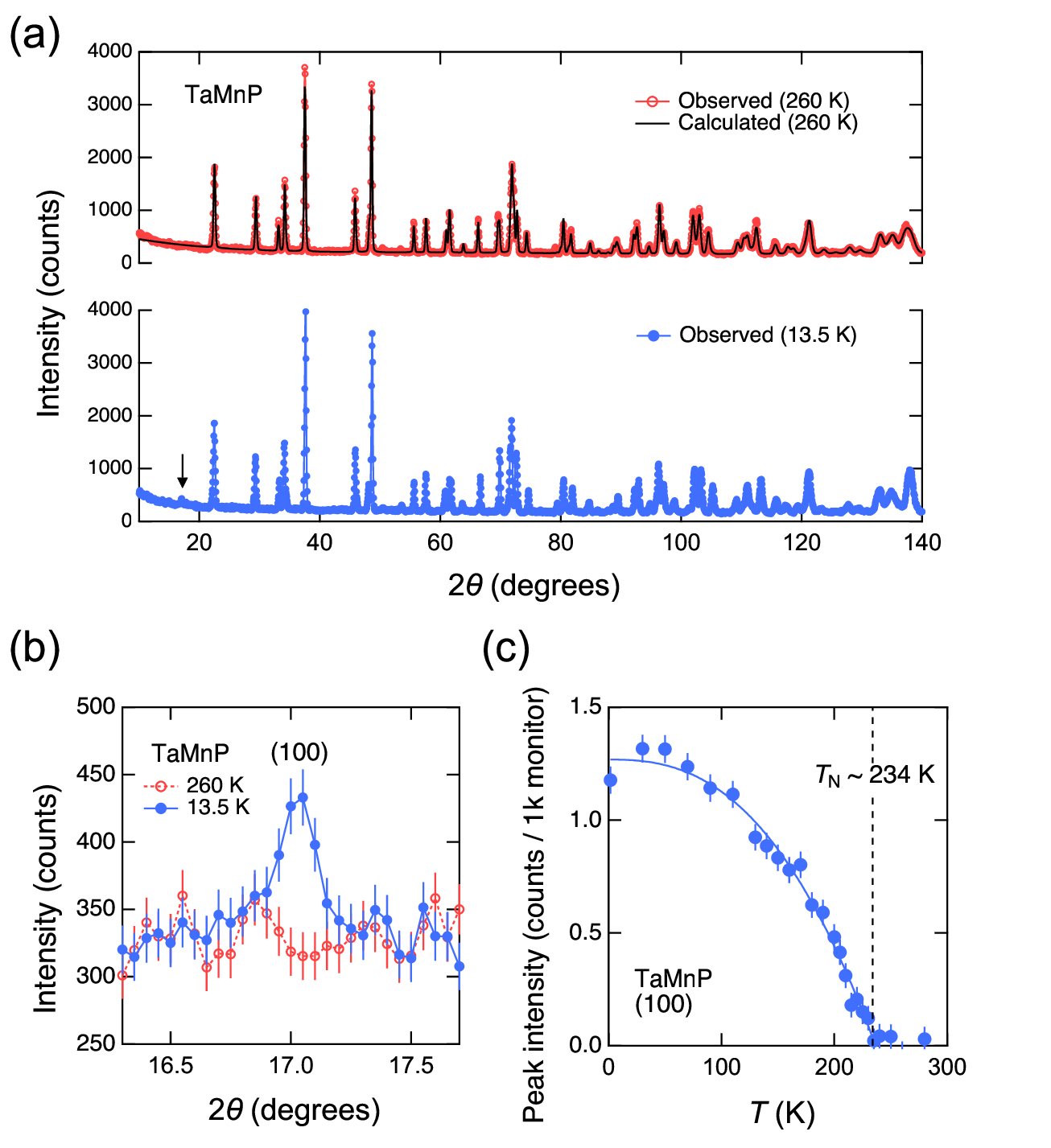}
\caption{(a) Powder neutron diffraction patterns of TaMnP at 260 and 13.5 K. (b) The 100 Bragg peak appears below $T_{\rm N}$. (c) Temperature dependence of the 100 Bragg peak intensity. The solid line is a guide to the eye. The strongest 100 peak and the absence of additional superlattice reflections indicates that the magnetic propagation vector is $q=0$ and the magnetic structure is represented by $A_u$($\Gamma_2$), $B_{3g}$($\Gamma_7$), or $B_{3u}$($\Gamma_8$) IRs.}
\end{figure}

\begin{figure*}[htb]
\includegraphics[width=15cm]{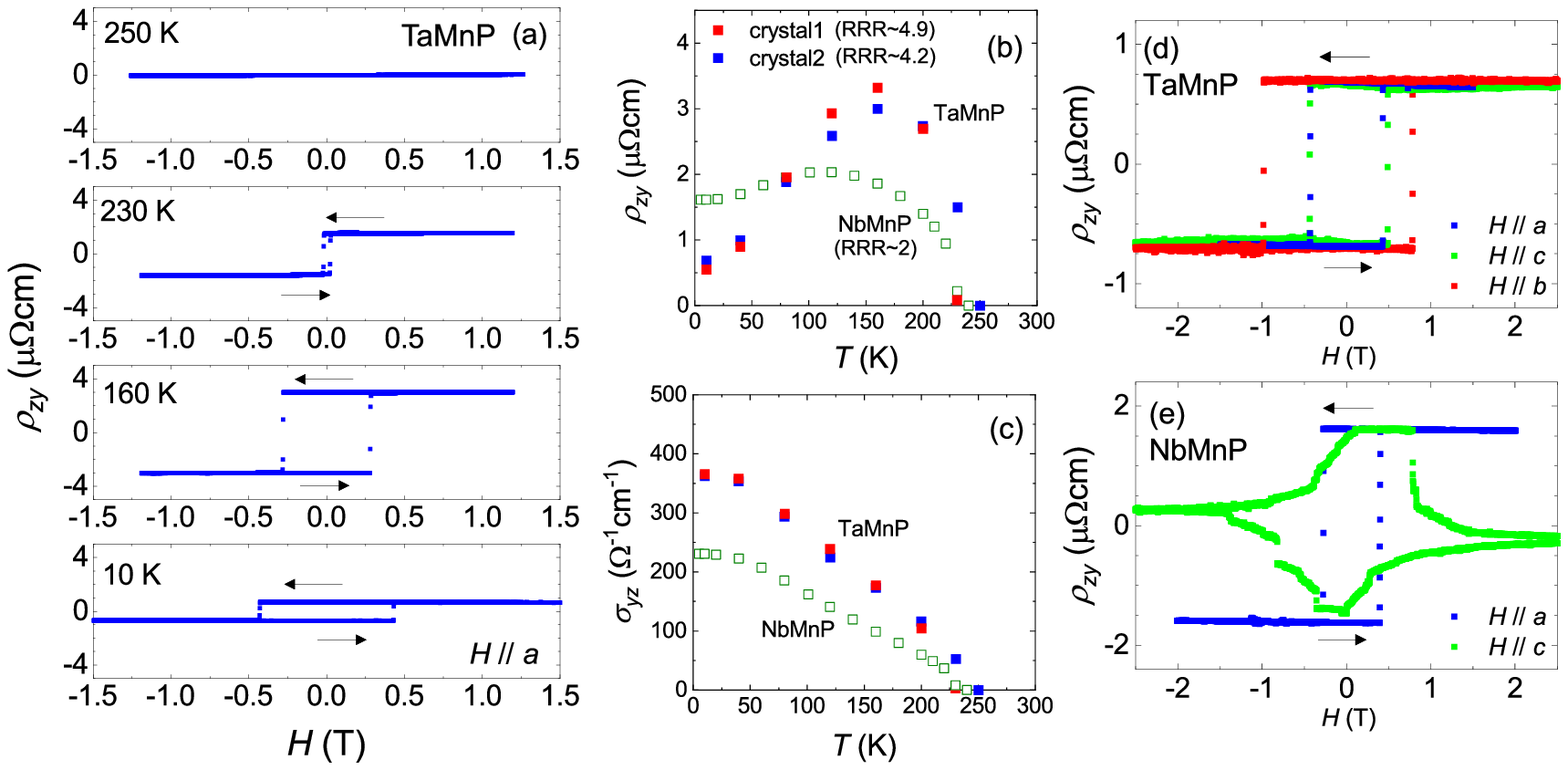}
\caption{(a) Hall resistivity $\rho_{zy}$ for TaMnP (Crystal 2) under $H \parallel a$. Significant AHE was observed in the magnetically ordered state, drawing a hysteresis loop against the positive and negative magnetic fields. The magnetic field along the $a$-axis abruptly switches the magnetic domain connected to the sign of the Hall effect. (b,c) Temperature dependence of $\rho_{zy}$ and $\sigma_{yz}$ estimated at zero fields. Futher, $\sigma_{yz}$ increases continuously toward low temperatures, reaching the maximum value of $\sigma_{yz} \sim 360-370$ $\Omega^{-1}$cm$^{-1}$. (d,e) Switching of $\rho_{zy}$ in different field directions for NbMnP \cite{Kotegawa_NbMnP} and TaMnP. In TaMnP, any field direction switches $\rho_{zy}$, i.e., AF domains of $B_{3g}$.}
\end{figure*}

The magnetic structure of TaMnP was investigated by neutron powder diffraction.
Figure 3(a) shows the diffraction patterns obtained at 260 and 13.5 K.
Above $T_{\rm N}$, the observed pattern was reproduced well by calculation based on the reported crystal structure with $Pnma$ symmetry.
After cooling to below $T_{\rm N}$, a tiny peak indicated by an arrow around 17$^{\circ}$ emerged and corresponded to (100), as magnified in Fig.~3(b).
The 100 peak became weaker with the increasing temperature, disappearing around $T_{\rm N}$ (Fig.~3(c)).
This ensured that the observed peak stemmed from the magnetic ordering.
The absence of any additional superlattice reflection indicated that the magnetic propagation vector is $q=0$, as in NbMnP\cite{Matsuda}.
In NbMnP, the 110, 012, and 300 Bragg peaks were observed in addition to 100, although they were invisible in TaMnP.
This is because a slight difference in the lattice constant resulted in an overlap with the strong 102 nuclear peak for 110, whereas weak signals could be beyond the sensitivity for the 012 and 300 reflections.
Table II presents the IRs allowed in the $q=0$ magnetic structure in the $Pnma$ space group.
Although eight representations were allowed, the strongest intensity for the 100 peak restricted them to three candidates: $A_u$($\Gamma_2$), $B_{3g}$($\Gamma_7$), and $B_{3u}$($\Gamma_8$). 
Among them, the net magnetization was only allowed in $B_{3g}$ along the $a$-axis.
The larger net magnetization along the $ac$ plane confirmed that the AF state of TaMnP was mainly represented by $B_{3g}$.
The AF structure represented by $B_{3g}$ is shown in Fig.~1.
The two diagonal Mn atoms, which are connected by the inversion symmetry, exhibit FM coupling, indicating that this magnetic structure exhibits $\mathcal{P}$ symmetry but not $\mathcal{PT}$ symmetry. 
The AF coupling between two Mn atoms is connected by screw operation, as shown in Fig. 1; therefore, the complete cancelation of the two magnetic moments is not guaranteed.
When the MPG (or IR) of the AF structure allows an FM state, it yields net magnetization through the DM interaction \cite{Kotegawa_NbMnP,Arai}; therefore, the AF structure in the $B_{3g}$ symmetry induces the FM component along the $a$-axis.
The small net magnetization along the $b$-axis was also experimentally observed in TaMnP (Fig.~2(d)).
This net magnetization has not been observed in NbMnP irrespective of the sample quality \cite{Zhao,Arai}.
The FM component along the $b$-axis corresponds to the $B_{2g}$ representation, which is irrelevant to the main component of the magnetic structure observed in the neutron scattering experiments.
Notably, the $B_{2g}$ representation allows only the FM component (Table II).
Therefore, the origin of the small net magnetization along the $b$-axis remains unclear.
The deficiency of the Mn site, as observed only in TaMnP, might have accounted for the emergence of this net magnetization; however, further investigation is required to clarify how it is induced.

\subsection{Anomalous Hall Effect and Unusual Domain Switching }

\begin{figure}[b]
\includegraphics[width=8cm]{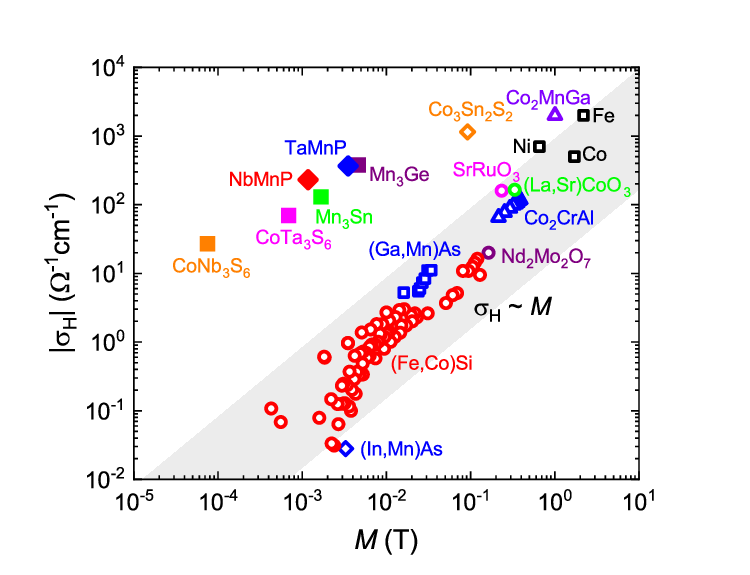}
\caption{AHC vs. net magnetization for various ferromagnets (open symbols) \cite{Manyala2004,Husmann2006,Taguchi2001,Mathieu2004,Onose2006,Miyasato2007,Sakai2018,Liu2018} and AF materials (closed symbols) \cite{Nakatsuji2015,Kiyohara16,Nayak16,Ghimire18,Park22,Kotegawa_NbMnP}. The rough relationship of $|\sigma_H| \sim M$ has been experimentally obtained for ferromagnets, as shown in the gray area. Evidently, AHC in the AF materials emerges irrespective of their small net magnetization. Here, we treated the magnetization for NbMnP as twice that of the experimental value since the measurement was performed for $H \parallel ac$, whereas the spontaneous magnetization was assumed to emerge along the $a$-axis.}
\end{figure}

\begin{figure*}[htb]
\includegraphics[width=16cm]{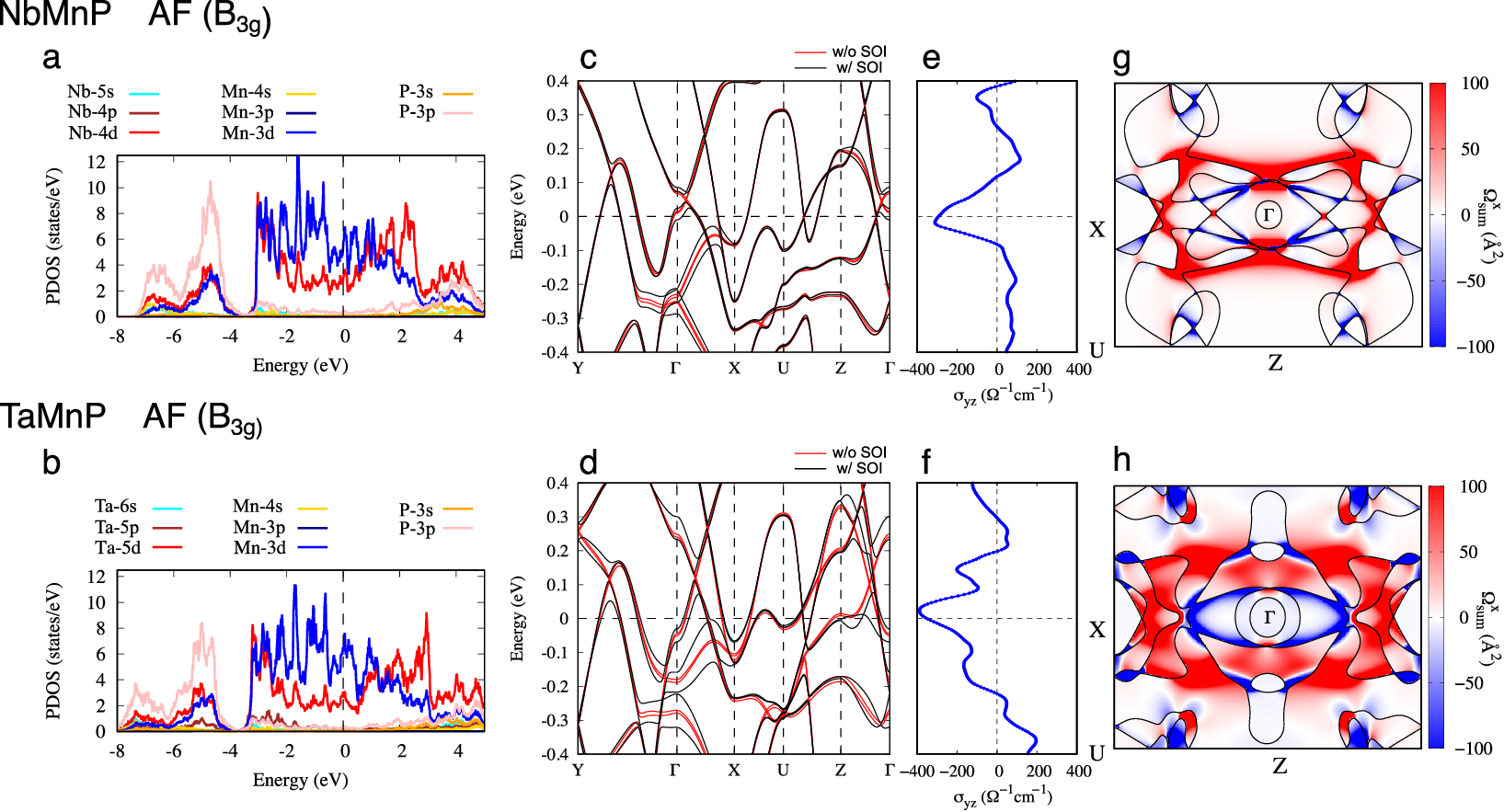}
\caption{Theoretical calculation of the electronic band structures of NbMnP and TaMnP. For the band dispersion, the black (red) curves show the calculation with (without) SOIs. The AF state is treated as $B_{3g}$. (g) and (h) represent the distribution of the Berry curvatures in the $k_y=0$ plane. The large Berry curvatures are distributed over the momentum space in TaMnP owing to the stronger SOI therein. The chemical-potential dependence of AHC is shown in the right panel for (e) NbMnP and (f) TaMnP. The calculated $\sigma_{yz} \sim 353$ $\Omega^{-1}$cm$^{-1}$ for TaMnP is fairly consistent with the experimental results.}
\end{figure*}

The presence of the $B_{3g}$ component in the magnetic structure was also confirmed by observing AHE.
Figure 4(a) shows the field dependence of the Hall resistivity $\rho_{zy}$, where the AHE is triggered by $B_{3g}$.
Here, magnetic fields were applied along the $a$-axis to align the AF domains.
A clear hysteresis against the positive and negative magnetic fields appeared in the magnetically ordered state.
The $\rho_{zy}$ exhibited weak field dependence except for the abrupt switching of its sign, and the nonzero $\rho_{zy}$ remained even at zero fields.
These findings confirmed that the observed Hall effect was dominated by AHE stemming from the magnetically ordered state.
Figure~4(b) shows the zero-field values in $\rho_{zy}$, whereas Fig.~4(c) shows its $\sigma_{yz}$, which was converted through $\sigma_{yz} \simeq \rho_{zy}/\rho_{yy}^2$, together with those of NbMnP \cite{Kotegawa_NbMnP}.
The $\rho_{zy}$ in TaMnP exhibited a maximum at approximately 160 K, and it was suppressed significantly toward low temperatures, as its RRR exceeded that in NbMnP.
The crucial quantity is $\sigma_{yz}$, which is determined by the band structure. This quantity increases at low temperatures, reaching a maximum ($\sim360-370$ $\Omega^{-1}$cm$^{-1}$), which is higher than that in NbMnP ($\sim230$ $\Omega^{-1}$cm$^{-1}$).

Qualitative difference between TaMnP and NbMnP appeared in the domain switching under the magnetic fields.
Figures 4(d) and (e) show $\rho_{zy}$ in different magnetic-field directions for the two systems.
In NbMnP, $\rho_{zy}$ was canceled under $H \parallel c$, indicating that $H \parallel a$ was essential for aligning the AF domain and generating the Hall response with the same sign.
These findings were consistent with that the magnetic structure contained the $B_{3g}$ symmetry, which allowed net magnetization along the $a$-axis.
In sharp contrast, TaMnP demonstrated clear switching of $\rho_{zy}$ in all the magnetic-field directions.
The similar switching fields for three axes could not be explained by the misalignment of the crystal.
The observation strongly indicated that the small net magnetizations along the $b$- and $c$-axes also coupled with the AF structure of the $B_{3g}$ representation.
They are symmetrically different, but the coupling among them unexpectedly emerged, imparting TaMnP with flexible controllability of the AF domains.
Such domain switching has been recently reported in LiNiPO$_4$ \cite{Kimura}, demonstrating an effective method for controlling AF domains.

The magnitudes of AHC of TaMnP and NbMnP were compared with those of other systems, including ferromagnets (Fig.~5), where AHCs were plotted against net magnetization in Tesla units.
The empirical linear relationship of $|\sigma_{H}| \sim M$ has been reported for various FM systems \cite{Manyala2004,Husmann2006}.
As shown in Fig.~5, itinerant magnets, oxides, and magnetic semiconductors are almost included in the gray area, with some exceptions in the Weyl systems exhibiting extremely large AHC \cite{Manyala2004,Husmann2006,Taguchi2001,Mathieu2004,Onose2006,Miyasato2007,Sakai2018,Liu2018}.
The tendency of $|\sigma_{H}| \sim M$ is obviously inapplicable to AF materials \cite{Chen2021}.
TaMnP is at a similar position to Mn$_3$Ge, an AF material exhibiting the highest AHC.

\subsection{Calculation of the Anomalous Hall Conductivity}

An inspection of the electronic structures is crucial to determining the emergence of large AHC was obtained in TaMnP.
Figures 6(a-d) show a comparison of the calculated density of states (DOS) and band structures of NbMnP and TaMnP.
For comparison, we set the $B_{3g}$ AF state for both materials.
Notably, the experimental AF structure of NbMnP was $B_{3g}+B_{2u}$ \cite{Matsuda}, while that of TaMnP was likely $B_{3g}$; this has not fully been confirmed. 
Our results indicated that the calculated magnetic moments of NbMnP and TaMnP were 1.47 $\mu_{\rm B}$ and 1.27 $\mu_{\rm B}$, respectively.
In addition to the majority of Mn-$3d$ observed near the Fermi level, the contribution of Nb-$4d$ or Ta-$5d$ to DOS was significantly large.
As shown in Figs.~6(c) and (d), the difference between the red and black curves indicates band modification induced by spin-orbit interaction (SOI).
The influence of SOI was more remarkable in TaMnP than in NbMnP owing to the difference between the Nb-$4d$ and Ta-$5d$ electrons.
The distributions of the Berry curvatures in the momentum space for NbMnP and TaMnP are shown in Figs.~6(g) and 6(h), respectively.
The values for NbMnP were slightly modified from the previous calculation for $B_{3g}+B_{2u}$ \cite{Kotegawa_NbMnP}; however, the tendency was similar, i.e., large Berry curvatures were obtained in the middle of the $\Gamma-Z$ direction.
The calculated AHC for $B_{3g}$was 293 $\Omega^{-1}$cm$^{-1}$, which was similar to the previously obtained value (276 $\Omega^{-1}$cm$^{-1}$).
For TaMnP, the large Berry curvatures were distributed over the momentum space, in $\Gamma-Z$ as well as $\Gamma-X$. 
Then, we obtained 353 $\Omega^{-1}$cm$^{-1}$, which was fairly consistent with the experimental results.
The above analysis confirmed that SOI of the Ta-$5d$ electrons significantly contributes to enhancing the Berry curvatures over the wide momentum space. The chemical-potential dependence of AHC (Fig.~6(f)) indicated that the imbalanced distribution of large Berry curvatures was realized particularly near the Fermi level, indicating the presence of a notable band topology, and this will be assessed in the future \cite{Huyen}.

\section{Conclusion}

In this study, we successfully synthesized TaMnP single crystals, whose physical properties were previously unknown.
We confirmed that these crystals exhibited AF transition at $\sim220-240$ K, which is accompanied by a small net magnetization ($\sim10^{-2} \mu_{\rm B}$/Mn).
The magnetic Bragg peak observed in the neutron scattering experiments indicated that the magnetic propagation vector is $q=0$.
The observation of significant AHE in $\rho_{zy}$ indicated that the AF structure exhibited a $B_{3g}$ representation, which is consistent with the results of the neutron scattering experiments.
AHC in TaMnP reached $\sim360-370$ $\Omega^{-1}$cm$^{-1}$, which is among the largest values reported for AF materials.
The theoretical calculations based on the $B_{3g}$ AF structure reproduced this large AHC, indicating that the large AHC in TaMnP was synergistically achieved by the Mn-$3d$ and Ta-$5d$ electrons, which were responsible for breaking the symmetry and facilitating strong SOI, respectively. 

Interestingly, we observed that the sign of $\rho_{zy}$ was switched by magnetic fields along all the crystal axes.
This confirmed that the AF structure of $B_{3g}$ coupled with small net magnetizations, whose symmetries differed form $B_{3g}$.
Our findings revealed a novel perspective for obtaining large Berry curvatures and flexible domain controllability in functional AF materials for future applications, including AF spintronics.

\section*{Acknowledgments}
We are grateful for the valuable discussions by Hisatomo Harima. This study was supported by the JSPS
KAKENHI Grant Nos. 18H04320, 18H04321, 19H01842, 21H01789, 21H04437, 21K03446, 23H01130, 23H04867, and 23H04871, Iketani Science and Technology Foundation, Hyogo Science and Technology Association, and Murata Science Foundation.
The numerical calculations were performed using MASAMUNE-IMR at the Center for Computational Materials Science, Institute for Materials Research, Tohoku University.

\end{document}